\documentclass[preview]{elsarticle} 

\usepackage{hyperref}

\journal{Nucl. Instr. and Meth. in Physics Research Sec. A}
\usepackage[binary-units = true]{siunitx}    
\DeclareSIUnit\neutron{neutron}
\DeclareSIUnit\Bq{Bq}
\DeclareSIUnit\uranium{U}
\DeclareSIUnit\n{n}
\usepackage{tikz}
\usetikzlibrary{backgrounds,positioning,fit,decorations.pathmorphing,arrows,shapes,calc,shadows,fadings}

\usepackage{xfrac} 
\usepackage[utf8]{inputenc}
\newcommand{\keepcomment}{0} 
\newcommand\hanno[1]{%
  \ifnum\keepcomment=1
  {\color{orange}{\textbf{Hanno: #1}}}
  \else
  \fi
}

\newcommand\linus[1]{%
  \ifnum\keepcomment=1
  {\color{blue}{\textbf{Linus: #1}}}
  \else
  \fi
}

\newcommand\bsg[0]{EJ-426/BSG
}

\newcommand\bsgns[0]{
EJ-426/BSG}

\newcommand\qrz[0]{EJ-426/QRZ
}
\newcommand\qrzns[0]{
EJ-426/QRZ}

\newcommand\bare[0]{%
PMT/BSG
}

\newcommand\barens[0]{
PMT/BSG}

\begin{document}
\sisetup{range-phrase=-,range-units=single,product-units = power}

\begin{frontmatter}

\title{Evaluation of the \emph{in-situ} Performance of Neutron Detectors based on EJ-426 Scintillator Screens for Spent Fuel Characterization}


\author[mymainaddress]{Hanno Perrey\corref{mycorrespondingauthor}}
\ead{hanno.perrey@nuclear.lu.se}
\author[mysecondaryaddress]{Linus Ros}
\author[mymainaddress]{Mikael Elfman}
\author[myfourthaddress]{Ulrika Bäckström}
\author[mymainaddress]{Per Kristiansson}
\author[mymainaddress,mythirdaddress]{Anders Sjöland}

\cortext[mycorrespondingauthor]{Corresponding author. Telephone:  +46 46 222 7708; Fax:  +46 46 222 4709}

\address[mymainaddress]{Division of Nuclear Physics, Department of Physics, Lund University, P.O. Box 118, SE-221 00 Lund, Sweden}
\address[mysecondaryaddress]{DVel AB, Scheelevägen 32, SE-223 63 Lund, Sweden}
\address[mythirdaddress]{Swedish Nuclear Fuel and Waste Management Company, SKB, P.O. Box 3091, SE-169 03  Solna, Sweden}
\address[myfourthaddress]{Vattenfall AB, SE-169 92 Stockholm, Sweden}

\begin{abstract}
The reliable detection of neutrons in a harsh gamma-ray environment is an important aspect of establishing non-destructive methods for the characterization of spent nuclear fuel.
In this study, we present results from extended in-situ monitoring of detector systems consisting of commercially available components: EJ-426, a $^{6}$Li-enriched solid-state scintillator material sensitive to thermal neutrons, and two different types of Hamamatsu photomultiplier tubes (PMT).
Over the period of eight months, these detectors were operated in close vicinity to spent nuclear fuel stored at the interim storage facility CLAB, Oskarshamn, Sweden.
At the measurement position the detectors were continuously exposed to an estimated (moderated) neutron flux of approx. $\SI{280}{\n\per\second.\cm\squared}$ and a gamma-ray dose rate of approximately \SI{6}{Sv/h}.

Using offline software algorithms, neutron pulses were identified and characterized in the data.
Over the entire investigated dose range of up to \SI{35}{\kilo\gray}, the detector systems were functioning and were delivering detectable neutron signals.
Their performance as measured by the number of identified neutrons degrades down to about 30\% of the initial value.
Investigations of the irradiated components suggest that this degradation is a result of reduced optical transparency of the involved materials as well as a reduction of PMT gain due to the continuous high currents.
Increasing the gain of the PMT through step-ups of the applied high voltage allowed to partially compensate for this loss in detection sensitivity  even when the detectors were highly irradiated.

The integrated neutron fluence during the measurement was experimentally verified to be in the order of $\SI{5e9}{\n\per\cm\squared}$.
The results were interpreted with the help of MCNP6.2 simulations of the setup and the neutron flux.

\end{abstract}



\end{frontmatter}



%
%
\section{Introduction}

The Swedish National Plan for Radioactive Waste~\cite{nationellPlan21, skb} outlines the handling of nuclear waste from Sweden's nuclear program.
Spent nuclear fuel from Sweden's reactors is to be encapsulated and stored in a final repository within the country.
The Swedish Nuclear Fuel and Waste Management Company (SKB)~\cite{skb} is responsible for managing the spent fuel and implementing the final geological repository.
SKB has developed the so-called KBS-3 concept~\cite{kbs3repository10}, which is a multi-barrier system based on encapsulating the spent fuel in copper canisters, embedding these in Bentonite clay and depositing them about 500 meters deep in the Swedish granitic bedrock.
It is necessary both from a nuclear safeguards verification and operational point of view to determine and verify key properties of each of the fuel assemblies prior to encapsulation, ideally via non-destructive assay~\cite{NDA} methods.
Among these parameters, the sub-critical neutron multiplicity of the spent fuel bundle is of particular importance.
Here, the neutron multiplicity is determined after the fuel has been removed from the reactor and is stored in water basins.
One method for measuring this observable is via coincident detection of pairs of neutrons emitted from decays within the spent fuel~\cite{NDA,DDSI}.
Due to the necessary proximity to the fuel assembly, this technique puts requirements on the employed detector systems.
In the foreseen measurement configuration, the detectors are expected to receive doses of up to \SI{10}{\gray\per\hour} and are expected to be exposed to roughly a million more gamma-rays than neutrons.
In this harsh radiation environment, the detectors need to be able to reliably identify neutrons with sufficient efficiency against the intense gamma-ray background and persistently handle the high accumulated doses.

SKB runs a long-term project with the goal of characterizing spent fuel assemblies stored at the Central Interim Storage Facility for Spent Nuclear Fuel (CLAB)~\cite{ClabWebPage}.
As part of this project, candidate detector technologies for the characterization setup have been operated under realistic conditions and their ability to identify neutrons has been monitored under a period of eight months.
The detector systems were built from commercially available materials and components.
The neutron converter material chosen for this study was the EJ-426-scintillator screen~\cite{EJ426Web}.
To distinguish the radiation-induced degradation of the scintillator from  the degradation of the entrance window of the photo-multiplier tube (PMT), three different PMTs were selected: one Silica glass PMT with EJ-426, one Borosilicate glass PMT with EJ-426 and one Borosilicate PMT without any scintillator.
Detector materials suitable for measurements in such an environment had been selected in a prior study performed at the Source Testing Facility (STF)~\cite{MessiSTF2017} at the Division of Nuclear Physics, Lund University~\cite{DetectorEval}.
In the selection process, the focus had been on the discrimination capabilities of scintillator materials between thermal neutrons and gamma-rays under moderate to high gamma-ray fluxes ($10^4$~--~$10^8$ gamma-rays per second in the detector).
ZnS(Ag), the scintillating component of EJ-426, is known to be suited for measurements of neutrons in similarly intense gamma-ray backgrounds\cite{wolfertz20_first_tests_gamma_blind_fast}.

Other authors have conducted dedicated studies on the radiation hardness of some of the employed components.
When the performance of ZnS(Ag) was evaluated after irradiation with \SI{14}{\MeV} neutrons up to fluences of $\SI{1e15}{\n/\cm^2}$, three radiation-induced effects were reported by~\cite{iida90_effec_neutr_irrad_optic_compon_fusion_diagn}:
A decrease in light output, a degradation of the energy resolution and an increase in noise levels.
In PMTs, two irradiation induced effects were found: degradation of gain and increase in the noise level.
For both the ZnS(Ag) and the PMTs, the onset of significant effects was after very high neutron fluences of $\SI{1e14}{\n/\cm^2}$ or more.
A change in the optical properties of the window material of the PMTs was identified to be one of the main causes to the degradation of the gain.
The window material has been shown to be decisive for the radiation tolerance of PMTs to neutrons as well as gamma-rays as reported by the manufacturer~\cite{HamamatsuPMTHandbook4E}.
The degradation in relative transmission in the entrance window when irradiated with gamma-rays is less than five \% for silica glass at the maximum measured dose of \SI{3.9e5}{\gray} while for borosilicate window effects of  $\SIrange{30}{60}{\%}$ were found already at a gamma-ray dose of \SI{1.1e4}{\gray}.

The study presented here has been performed at CLAB, where the spent fuel from Sweden's nuclear power plants is stored in water-filled basins.
CLAB thus provided a realistic test bed for the long-term performance of the complete detector system in a high-intensity mixed-field radiation environment.
However, the facility's mission is primarily to handle and store the spent fuel safely and not the support of research projects.
This resulted in somewhat limited access to the measurement system during the campaign and requirements on the ability of the system to autonomously handle environmental changes such as power outage and changes in temperature or humidity.

The results presented here are expected to be an important input for design decisions within the project as well as for applications in other radiation-harsh environments where thermal neutrons are to be measured against a strong gamma-ray background for prolonged times.

\section{Neutron Detector Assemblies}
\label{sec:detectors}

\begin{figure}
    \centering
    \includegraphics[width=.5\textwidth]{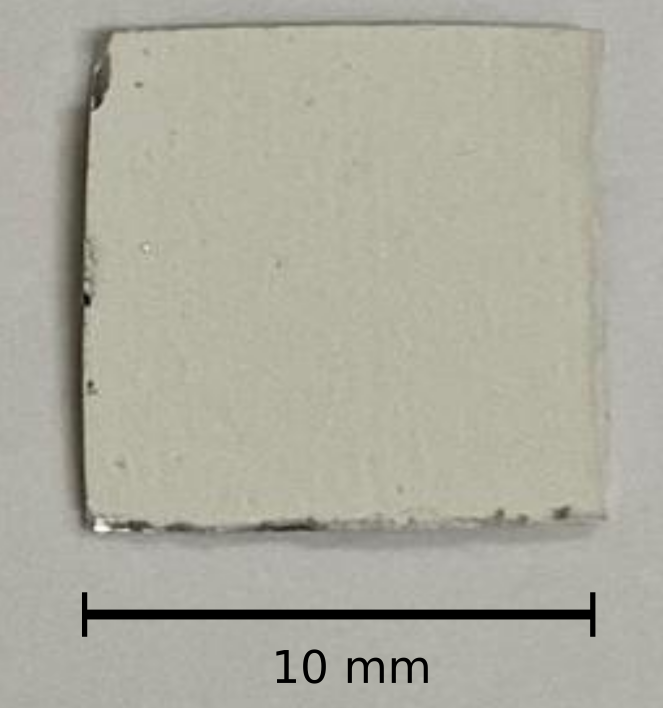}
    \caption{Photograph of a $\SI{1}{\cm\squared}$ piece of EJ-426 scintillator screen cut into a square shape.}
    \label{fig:Ej426}
\end{figure}

The neutron detectors employed in this study were assembled from a neutron scintillator material coupled to photo multiplier tubes.
Figure~\ref{fig:Ej426} shows a photograph of the EJ-426 thermal neutron scintillator screen used in this study.
The screen consists of a mixture fine grains of $^6$LiF and ZnS:Ag dispersed in a colorless binder and layered on a backing material.
The $^6$Li content of EJ-426 is enriched to at least 95\% of Li atoms~\cite{EJ426Web}.
The material provides neutron sensitivity via the capture reaction on the $^{6}$Li: The $^6$Li captures a neutron and subsequently decays to $^{3}$H and $^{4}$He fragments which share the released energy of $\SI{4.78}{\MeV}$.
These end products travel the scintillator and interact with the ZnS:Ag which results in scintillation light in the wavelength interval $\SI{400}{\nano\meter}$~--~$\SI{560}{\nano\meter}$ and a peak wavelength at $\SI{450}{\nano\meter}$.

The pulse shape of the resulting signal differs significantly depending on whether the pulse originates from gamma-ray absorption or neutron interaction in the EJ-426 scintillator.
Absorption of a gamma-ray produces a short pulse with a fast leading edge and a fast decay component, with a decay constant in the order of $\SI{3}{\nano\second}$.
The pulse from the charged reaction products of a neutron capture on the \textsuperscript{6}Li, on the other hand, produces a pulse with fast leading edge and a decay consisting of three components:
a fast, a semi-slow and a slow component with decay constants of $\SI{3}{\nano\second}$, $\SI{0.16}{\micro\second}$, and $\SI{4.2}{\micro\second}$, respectively~\cite{EJ-426Decay}.
The fast light emission is a result of bi-molecular electron-hole recombination while the light for the slow phosphorescence mostly originates from the thermal release of electrons from traps into the conduction band.
These slower components become dominant when the scintillator is exited by particles causing higher ionization density, e.g. $^{3}$H and $^{4}$He (neutron absorption) compared to the lower ionization densities from the secondary electrons released in gamma-ray absorption~\cite{EJ-426Decay}.

EJ-426 emits around 50000 photons/MeV~\cite{EJ-426-Optimisation} which is considered a bright scintillator material for thermal neutron detection~\cite{eijk12_inorg_scint_therm_neutr_detec}.
As the scintillator mixture is opaque to the particular wavelength emitted, practical shapes of the scintillator are limited to thin sheets or require coupling and read-out through wave-length shifting optical fibers as demonstrated by others~\cite{hildebrandt16_detec_therm_neutr_using_zns_ag}.
EJ-426 comes with different matrix thicknesses, $^6$Li to ZnS mass ratios, and backing materials~\cite{EJ426Web}.
The particular screen used in this setup is EJ-426-0 with a $^6$Li to ZnS mass ratio of 1:3 and a thickness of $\SI{0.32}{\mm}$ attached to a $\SI{50}{\micro\meter}$ thick aluminum backing.

$\SI{1}{\cm^{2}}$ pieces of EJ-426-0 were mounted onto two different types of photo multiplier tube~(PMT):
a $\SI{19}{\mm}$ diameter Hamamatsu R1450 PMT\footnote{This PMT and voltage divider is from own stock and has been exposed to a limited amount of radiation before the described measurement.}~\cite{R1450} with a E974-13 voltage divider~\cite{E974-13} and a 19 mm diameter Hamamatsu H6613~PMT assembly~\cite{H6613} incorporating a R2067~PMT~\cite{R2076}.

The R1450 and the R2076 have 10 and 8 dynode stages, respectively.
Both models have a Bialkali photo-cathode and a maximal spectral response at a wavelength of $\SI{420}{\nano\meter}$.
For the purpose of this study, the key difference between these two models is the window material:
While in the R1450 Borosilicate glass (BSG) has been employed, the R2076 features Quartz (QRZ) glass which is considered more radiation hard~\cite{H6613}.

\begin{figure}
    \centering
    \includegraphics[width=.5\textwidth]{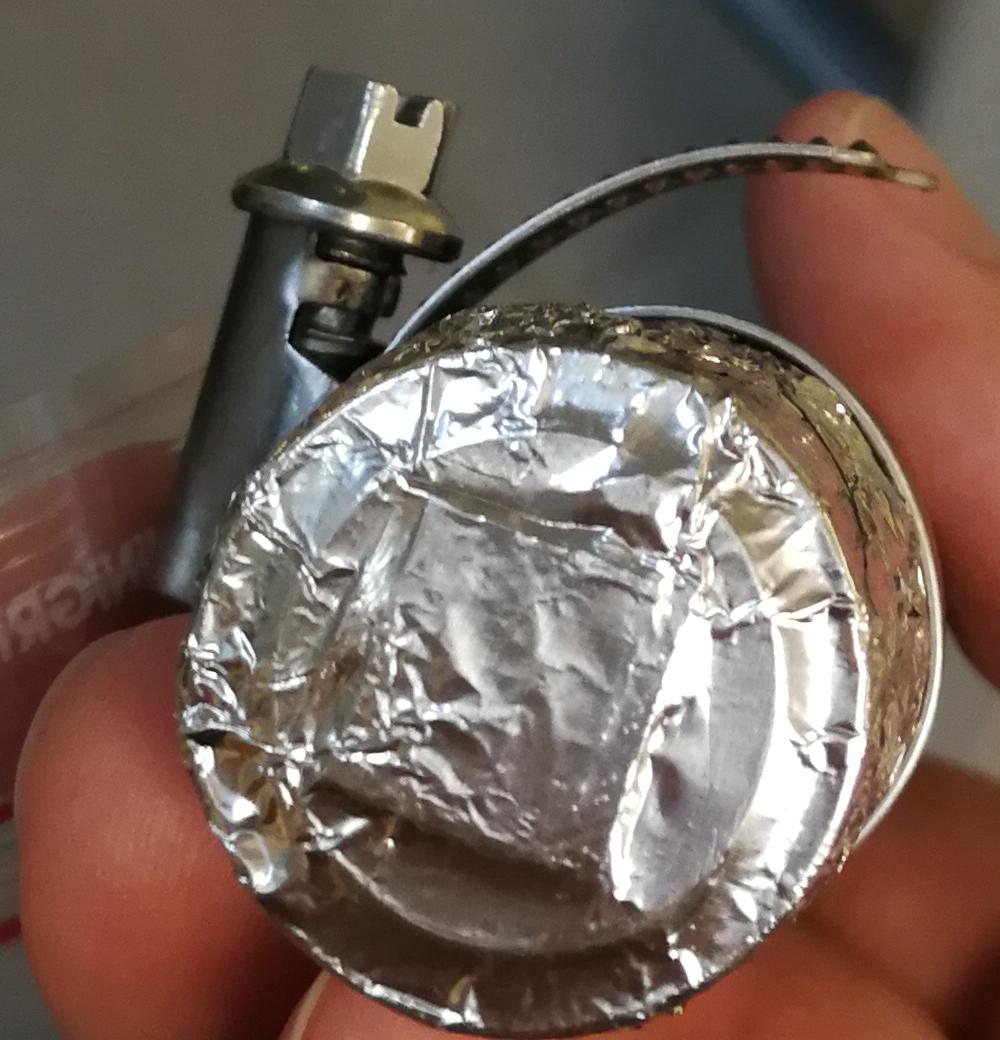}
    \caption{EJ-426 mounted on the Hamamatsu H6613 PMT assembly window using aluminum foil and a hose clamp as fixture. The materials were chosen to fulfill the radiation hardness requirements of the setup.}
    \label{fig:detector2}
\end{figure}

For mounting, the EJ-426 scintillator was pressed and held against the PMT window using aluminum foil and held by a hose clamp around the shaft of the PMT.
This particular procedure avoids the use of vinyl tape or optical grease, as the these products are not expected to withstand the radiation levels without significantly changing their properties.
A fully mounted detector can be seen in Figure~\ref{fig:detector2}.

Two detectors were assembled this way, one based on the R1450 and one based on the R2076.
These are referred to in the text as \bsg and \qrz detectors, respectively, emphasizing the type of PMT window being used.
In both the configurations of PMT and voltage divider given above, the maximum voltage the PMT assembly is specified for is $\SI{-1800}{\volt}$.

As a reference, an additional bare R1450 PMT and voltage divider were enclosed in aluminum foil using the recipe described above without mounting any scintillator screen.
This detector is used to monitor interactions of gamma-rays and neutron with the material of the PMT itself as well as to study the degradation of the PMT on its own.
In the text, the bare R1450 PMT is referred to as \bare.

\subsection{Detector vessel and commissioning}
\label{sec:detect-vess-comm}

\begin{figure}
    \centering
    \includegraphics[width=.5\textwidth]{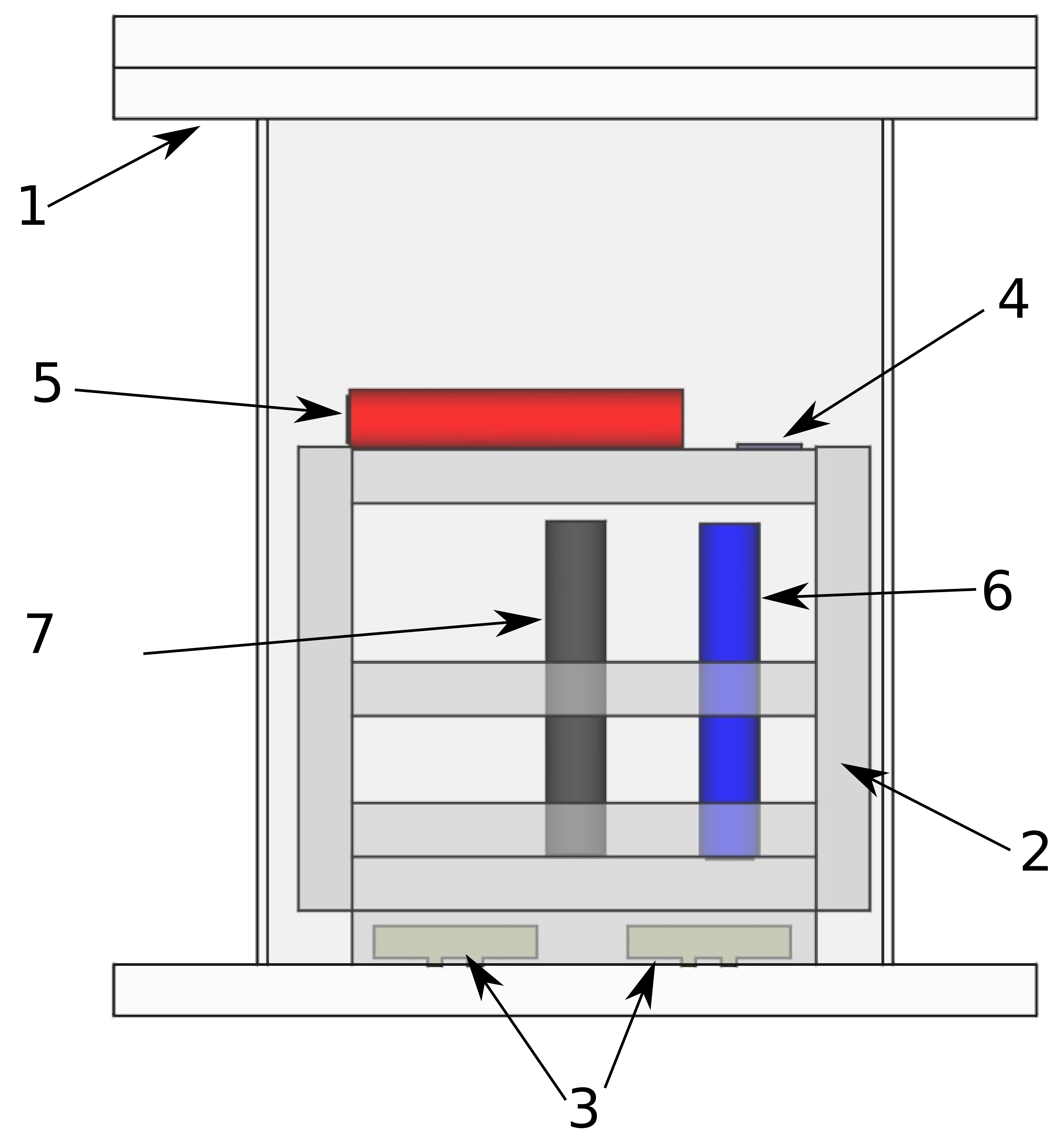}
    \caption{Equipment in the detector vessel. 1: Detector vessel, 2: High density polyethylene bricks, 3: Water sensors, 4: Tantalum foil, 5: Borosilicate Glass PMT (\barens), 6: Borosilicate Glass PMT with EJ-426 scintillator (\bsgns), 7: Silica Glass PMT with EJ-426 scintillator (\qrzns). See Sections~\ref{sec:detectors} and~\ref{sec:neutronDose} for details.}
    \label{fig:MeasurementContainerEq}
\end{figure}

All detector assemblies were mounted inside a rectangular portable, submersible and water-tight stainless steel (EN 1.4301) container.
This detector vessel had the external dimensions \SI{36x16x46}{\cm} and a removable top lid.
The lid was secured with 8 screws and sealed with EPDM 70~\cite{toman07_elastomer_handbook}.
A waterproof silicon (GP50) hose was attached and used as feed-through for cables.
Inside the detector vessel the detectors were mounted and insulated from the metal of the container with bricks of polyethylene plastic.
The bricks were machined to fit the box, keeping the detectors and electronics in position.
The polyethylene also functioned as moderator material for (fast) neutrons.
Underneath the plastic, two magnetic reed-switch water sensors by GRI~\cite{GRI} were placed to raise an alarm if water should leak into the detector vessel.
A schematic of the detector vessel indicating the detector positions is shown in Figure~\ref{fig:MeasurementContainerEq}.

\begin{figure}[htpb]
    \centering
    \includegraphics[keepaspectratio=true,width=\textwidth]{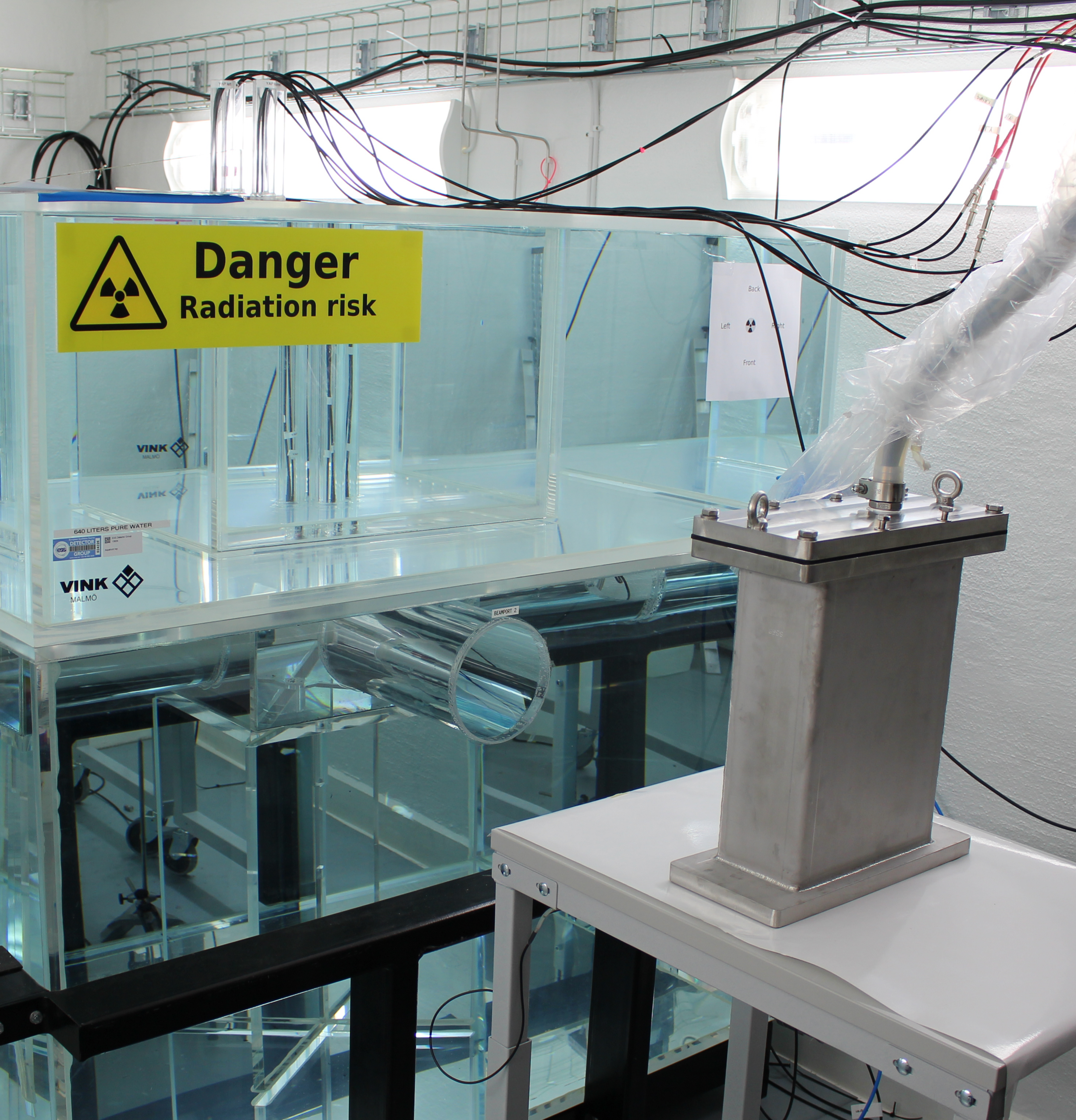}
    \caption{Photograph taken during the commissioning of the equipment at the STF at Lund University. Shown to the right is the detector vessel with attached silicon cable feed-through and protective plastic wrapping. Visible on the left is the STF's Aquarium, a water-filled tank providing beams of neutrons from a beryllium-based source used for the irradiation of the detector vessel during the tests.}
    \label{fig:MeasurementContainerPic}
\end{figure}

Before the installation on site, all detectors as well as the detector vessel were tested and commissioned at the STF at Lund University.
Figure~\ref{fig:MeasurementContainerPic} shows a photograph of the detector vessel taken during the commissioning work.
During these tests, the detectors were irradiated with neutrons from moderated beryllium-based sources.

\section{Facility and Experimental Setup}
\label{sec:facility}

The interim storage facility CLAB is located close to the Swedish city Oskarshamn and has been in operation since 1985.
After an initial cooling period at the reactor, spent nuclear fuel is transported to CLAB and will remain there until the future encapsulation plant CLINK (The Combined CLAB and encapsulation facilities) is operative~\cite{nationellPlan21}.
The facility is able to receive approximately \SI{300}{\tonne} uranium fuel from Swedish nuclear power plants each year and is presently permitted a total capacity of \SI{8000}{\tonne} uranium.
The fuel found at CLAB is from pressurized water reactors (PWR) or boiling water reactors (BWR), as both reactor types are employed in the Swedish nuclear industry.
During storage, the spent nuclear fuel is kept in \SI{13}{\meter} deep water basins located \SI{30}{\meter} under ground.

To study the long-term, in-situ performance of the neutron detector assemblies in the strong gamma-ray fields from the remaining activity of the fuel, permission was granted to install the detector vessel in one of the storage basins at CLAB for a period of eight months.

The detector vessel was submerged into the water between the concrete wall of the basin and a so-called fuel compact cassette containing 25~BWR fuel elements.
The fuel elements were arranged in a \SI{5x5}{} geometry with each element having approximate dimensions \SI{13x13x400}{\cm} (WxDxH).
All elements in the cassette had a cooling time of around 20 years and a burn-up of \SIrange{40.7}{44.4}{\giga\watt.\day/\tonne.\uranium}.

\begin{figure}
    \centering
    \includegraphics[width=.7\textwidth]{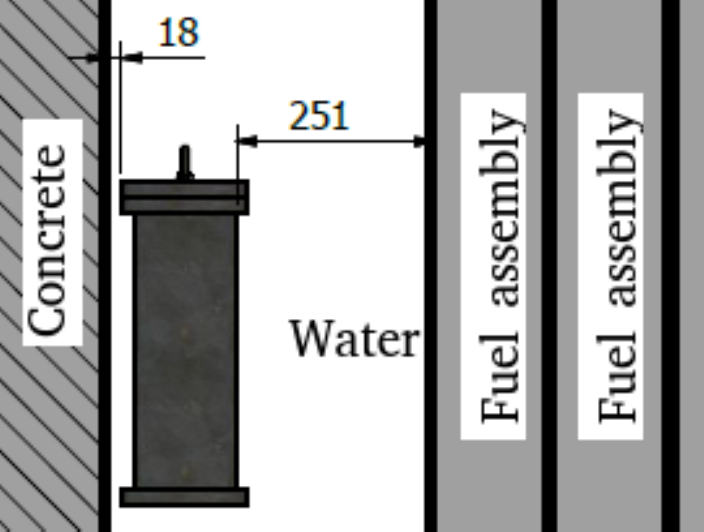}
    \caption{Drawing of the positioning of the detector vessel (left of center) in the storage pool between the concrete wall (left) and the spent nuclear fuel assembly (right). The distances specified in the drawing are in units of mm.}
    \label{fig:ContainerPos}
\end{figure}

The detector vessel  was fixed in position at a depth of \SI{10.5}{\meter} and \SI{2.5}{\meter} above the bottom of the pool.
In this position, it was facing the front of the fuel assembly at a distance of about \SI{25}{\cm} as indicated in figure~\ref{fig:ContainerPos}.
Between the concrete wall and the detector vessel, there was about~\SI{2}{\cm} of water.

\subsection{Neutron flux estimation}
\label{sec:neutronDose}
In order to experimentally verify the neutron flux, a high purity tantalum foil (\SI{99.999}{\%})~\cite{goodfellow} was mounted inside the detector vessel as indicated in Figure~\ref{fig:MeasurementContainerEq}.
The piece of tantalum had dimensions \SI{25x25x2}{\mm} and a weight of \SI{21.7}{\g}.
Neutrons interact with the tantalum mainly via the neutron capture reaction $^{181}\mathrm{Ta} (\mathrm{n},\gamma) \rightarrow ^{182}\mathrm{Ta}$.
The thus formed $^{182}$Ta decays, with a half-life of 114.61 days, via $\beta$\textsuperscript{-} decay to exited levels of $^{182}$W.

In order to relate the activity in the irradiated tantalum foil with the energy dependent neutron flux in the detector vessel, neutron activation simulations were performed using MCNP6.2~\cite{MCNP}~\cite{activation}.
All the simulated 25 BWR fuels are described by BWR-7, a MCNP input file, in the SKB-50 set of high fidelity models of spent nuclear fuel~\cite{osti_1326492}.

After the measurement campaign, the tantalum foil was recovered and the activity concentration, i.e. the amount of activated tantalum relative to its mass, was measured.
The equipment used consisted of a Caen DT5780 MCA and an Ortec HPGe detector calibrated for absolute efficiency.
The final activity concentration from $^{182}$Ta in the tantalum foil was calculated for six $\gamma$-lines~\cite{TantalumIrrad}.
Since the tantalum sample has a a non-negligible thickness, the activity was corrected for the self-absorption on the sample as determined from MCNP6.2 simulations.

\subsection{Gamma-ray dose measurements}
\label{sec:gammaDose}
\sisetup{exponent-to-prefix=true,range-units=repeat} 
During the installation the dose from gamma-rays was measured at several positions along the fuel elements using a RDS-31 survey meter and a GMP-12SD external gamma dose probe from Mirion~\cite{RDS-31, GMP-12}.
The probe is specified for an energy interval between $\SIrange{60}{6e3}{\keV}$ and dose rates between \SI{10e-6}{\sievert\per\hour}-\SI{10}{\sievert\per\hour}.
While lowering the probe along the fuel canister the measured dose rate varied between $\SIrange{4}{8}{\sievert\per\hour}$.
At the position of the detector vessel as well as \SI{0.5}{\m} above and below the dose rate was measured to be \SI{6}{\sievert\per\hour}.
We expect this value to be accurate within 20\% as it is affected by positional variations of the flux close the the spent fuel container as well as the measurement accuracy of the probe and survey meter.
\sisetup{exponent-to-prefix=false,range-units=single} 

\section{Data Acquisition and Data Analysis}

\begin{figure}[htbp]
  \centering
  \includegraphics[width=10cm]{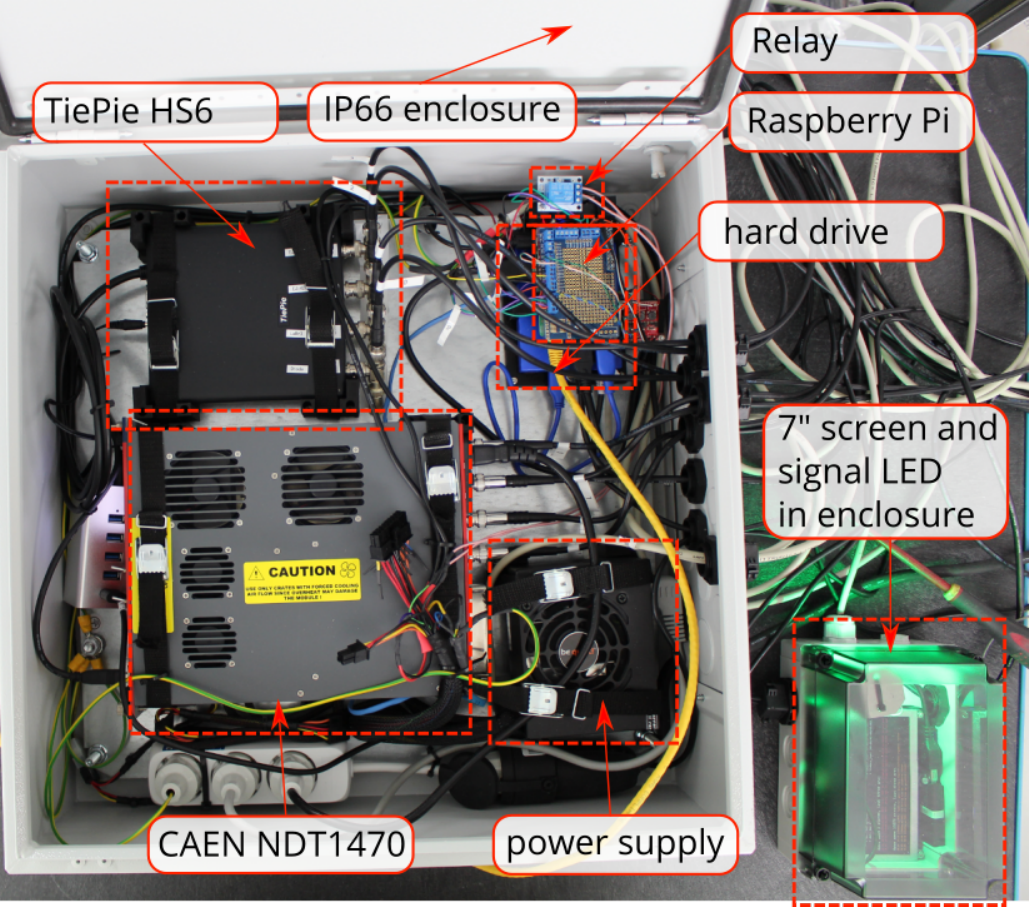}
  \caption[Picture of the data-acquisition hardware.]{A photograph of the data acquisition hardware and high voltage power supply inside the IP66-rated enclosure with labels identifying the various hardware components.}
  \label{fig:daqhw}
\end{figure}

During the measurement campaign, only sporadic access to the system and no remote connectivity were foreseen.
Autonomous operation of the data acquisition system (DAQ) was therefore a priority.
Due to the potentially high humidity of the environment at CLAB, the entire data acquisition system had to be placed in IP66-rated enclosure.
Figure~\ref{fig:daqhw} shows the open enclosure and identifies the various hardware components contained therein.

The signal outputs from the detectors were connected to a TiePie Handyscope HS6~DIFF-1000XMG digitizer.
The HS6 digitizer is a four-channel waveform digitizer with configurable resolution between 8-bit and 12-bit over the configurable voltage range, a sample frequency of up to $\SI{1}{\giga\hertz}$ and a bandwidth of up to $\SI{250}{\mega\hertz}$.
The on-board buffer capacity of the HS6 is $\SI{256}{\mega\byte}$ and allows to store $\sim$\SI{250}{\milli\second} of continuous samples in a single acquisition.
High voltage was supplied to the detectors by a four-channel Caen~NDT1470 module.

Both the HS6 digitizer and the HV module were controlled via USB interface by a Raspberry~Pi~3B+ computer chosen for its low power consumption.
The Raspberry Pi is equipped with a PiJuice module and a battery which together act as uninterruptible power system to protect the Raspberry Pi against possible (scheduled) power cuts at CLAB.

Control of the HS6 digitizer and HV, monitoring of sensors and data handling were performed by the Python-based DAQ software \texttt{clabdaq}~\cite{clabdaq_repo} running on the Raspberry Pi.

Copies of the recorded data and performance logs were stored on a thumb drive that could periodically be exchanged and shipped off-site for analysis.
Through the replacement thumb drive, modifications of the operational parameters and the software code of the data acquisition system could be performed.
With the exception of the physical exchange, no interaction with the system is required under normal operation.
The status of the system was indicated through a $\SI{7}{inch}$ LCD screen and a high-powered RGB-LED.
The latter was used to alert on-site personal in case of any failure that the system could not recover from.

\subsection{Recorded Data Samples and Offline Neutron Identification}
\label{sec:data-sampl-analys}

In the period from August 2020 to April 2021, the data presented in this study were recorded for each of the connected detectors on a daily basis\footnote{Data has been recorded successfully on 233 out of the 245 days of the measurement campaign, corresponding to an up-time of 95\%.}.
The time window for data taking has been set to the hours of the night to reduce the influence that routine operation at CLAB could have on the measurements.
For each day the start of data recording was randomly chosen within the time window.
Each capture was triggered via software to avoid introducing bias by a fixed trigger threshold.
For each of the detectors, the signals were sampled and read-out subsequently using a set of different digitizer configurations.
These varied mainly in the resolution and sample frequency.
Between the two configurations, 8-bit and 12-bit resolution with \SI{1}{\giga\hertz} and \SI{500}{\mega\hertz} sampling frequency, respectively, no significant differences were found for the purposes of this study.
The input voltage range was set to $\SI{0.2}{\volt}$.

A total of $\SI{1}{\second}$ of data was recorded for each detector per acquisition.
A total of 381 acquisitions were taken per detector of which 130 were measured during the initial 12 hours where the acquisition was triggered approximately every 5 minutes.

The data were analyzed offline using Python-based algorithms to identify neutron candidates in the digitized signals.

\begin{figure}
    \centering
    \includegraphics[width=\textwidth]{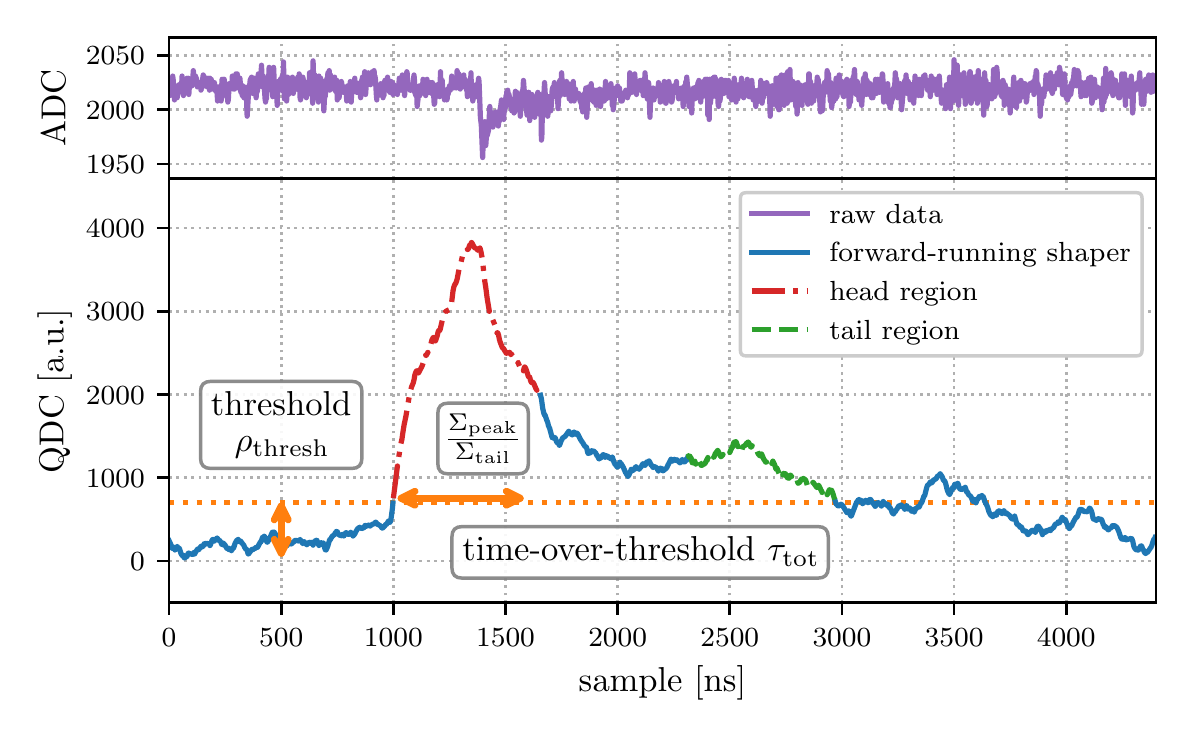}
    \caption{Example for a neutron candidate pulse and illustration of the neutron candidate identification algorithm: shown are a part of the raw 12-bit ADC data with a typical lower-amplitude neutron pulse ($\sim \SI{6}{\milli\volt}$) as recorded by the digitizer (top) and the resulting QDC values of the forward-running shaper algorithm (bottom). Threshold, time-over-threshold and head-versus-tail selection criteria applied to the QDC data are indicated in the figure. See text for details.}
    \label{fig:ncandidates}
\end{figure}

Figure \ref{fig:ncandidates} shows a typical neutron pulse and illustrates the algorithm to select neutron candidates.
The latter is tuned to the characteristic pulse-shape produced by neutrons reacting inside EJ-426.
First, the raw digitized signal is processed by a forward-running shaper which integrates the baseline-subtracted pulse over $\SI{0.4}{\micro\second}$.
For each sample point, this integration is performed by summing up the ADC values\footnote{Given the resolution and input voltage range, the ADC values of the digitizer can be trivially converted to Voltages. However, for easier data handling, the integer ADC values were used in the selection process.} of all subsequent samples within the integration window.
Of these shaped pulses only those are considered as neutron candidates which:
(a) are above a threshold $\rho_{\mathrm{thresh}}$;
(b) have a time-over-threshold (ToT) above a value $\tau_{\mathrm{tot}}$;
(c) have a pronounced peak in first \sfrac{1}{3} of the ToT region followed by a tail in the last \sfrac{1}{3} as parameterized by the ratio $\Sigma_{\mathrm{peak}}/\Sigma_{\mathrm{tail}}$ where $\Sigma_{\mathrm{peak}}$ and $\Sigma_{\mathrm{peak}}$ denote integrals over the respective regions of the distribution;
(d) are at least $\Delta \tau$ apart from each other (to avoid double-counting and overlapping pulses).

The value for $\rho_{\mathrm{thresh}}$ used was 100 and 700 for 8-bit and 12-bit data sets, respectively, $\tau_{\mathrm{tot}}$ was set to \SI{1.2}{\micro\second}, and $\Sigma_{\mathrm{peak}}/\Sigma_{\mathrm{tail}}$ to 1.7 and $\Delta \tau$ to \SI{5}{\micro\second}.
These parameters were tuned to find as many neutron candidates as possible while avoiding any mis-identification due to noise.
For selected runs, this was verified by manually inspecting the algorithmically identified pulses by eye.

For the two detectors \qrz and \bsgns, a total of 7093 and 2006 neutron candidates have been identified in the acquired data, respectively.

\section{Results and Discussion}

The detector systems were irradiated in the mixed-field environment up to a dose of approx.~\SI{35}{\kilo\gray}.
During the installation of the measurement setup at CLAB, the operational parameters of the detectors were determined and later continuously monitored over the entire dose range.
The performance of the detectors was evaluated in terms of the number of identified neutron candidates and their amplitudes as a function of the accumulated dose.
After the measurement campaign was concluded, the individual detector components were tested in the laboratory and compared to earlier reference measurements.

\subsection{Operational PMT parameters in a harsh radiation environment}
\label{sec:HV}

\begin{figure}
    \centering
    \includegraphics[width=\textwidth]{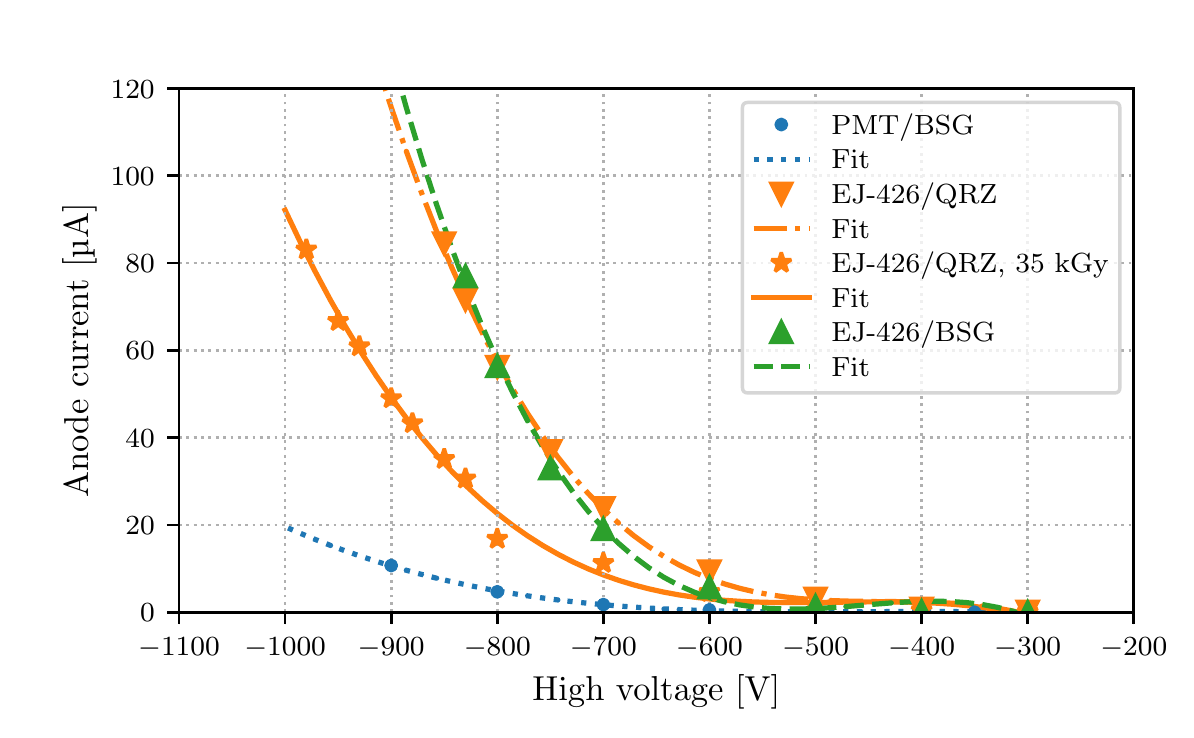}
    \caption{Measured anode current as a function of detector high voltage at the measurement position next to the spent fuel assembly. Markers indicates measured data for \bare (blue dots), \qrz (orange downward-pointing triangle and orange stars after irradiation with \SI{35}{\kilo\gray}) and \bsg (green upward-pointing triangles). The lines of the corresponding color indicate  third-degree polynomial fits to the respective data to guide the eye. (For interpretation of the references to color in this figure caption, the reader is referred to the web version of this article.)}
    \label{fig:CurrVsHV}
\end{figure}

As final part of the installation process at CLAB, the high voltage to the detectors was increased in small steps to determine their operational point while positioned in the final measurement position.
The same procedure was followed before recovering the detector vessel.
After each voltage increase the anode current from the PMTs was measured using an Agilent U1253B multi-meter, with a current resolution of $\SI{0.01}{\micro\ampere}$.
Additionally, the waveform of the current signal was captured with a LeCroy WaveJet 324 oscilloscope.
Figure~\ref{fig:CurrVsHV} shows the measured anode current as a function of the high voltage applied to the different detectors including data taken for \qrz after the detector had been exposed to \SI{35}{\kilo\gray}.
The lines are third-degree polynomial fits to the respective data set to guide the eye.
For the detectors with EJ-426 scintillator screen, the anode current increases swiftly.
Extrapolating the trend of the data, the anode current would reach the maximum value of \SI{100}{\micro\ampere} specified for this PMT type already at a voltage of around \SI{-870}{\volt}. %
As a reference, the anode current was measured to be \SI{0}{\micro\ampere} at a voltage of \SI{-900}{\volt} for all three detectors before the detector vessel was submerged into the pool.
In case of the detectors with installed scintillator, \bsg and \qrzns, the increased anode current in the radiation field at the measurement position is expected to be mainly a result of the high number of $\gamma$-rays interacting in the scintillator.
The resulting scintillation light produces a constant current through the PMTs.
In comparison, the \bare detector shows a slower rise in anode current as the high voltage increases which is likely originating from $\gamma$-ray interactions in the glass body or the dynodes of the PMT itself producing secondary electrons and thus a constant current.

All three PMTs were set to an initial high voltage of \SI{-800}{\volt}, in order to keep a sufficient margin before the maximum anode current is reached.

After irradiation, the gain curve of the \qrz is similar to the initial one but shifted towards more negative values.
This indicates either a reduction of the PMT gain, or less light from the scintillator entering the PMT, or both.

\begin{figure}
    \centering
    \includegraphics[width=\textwidth]{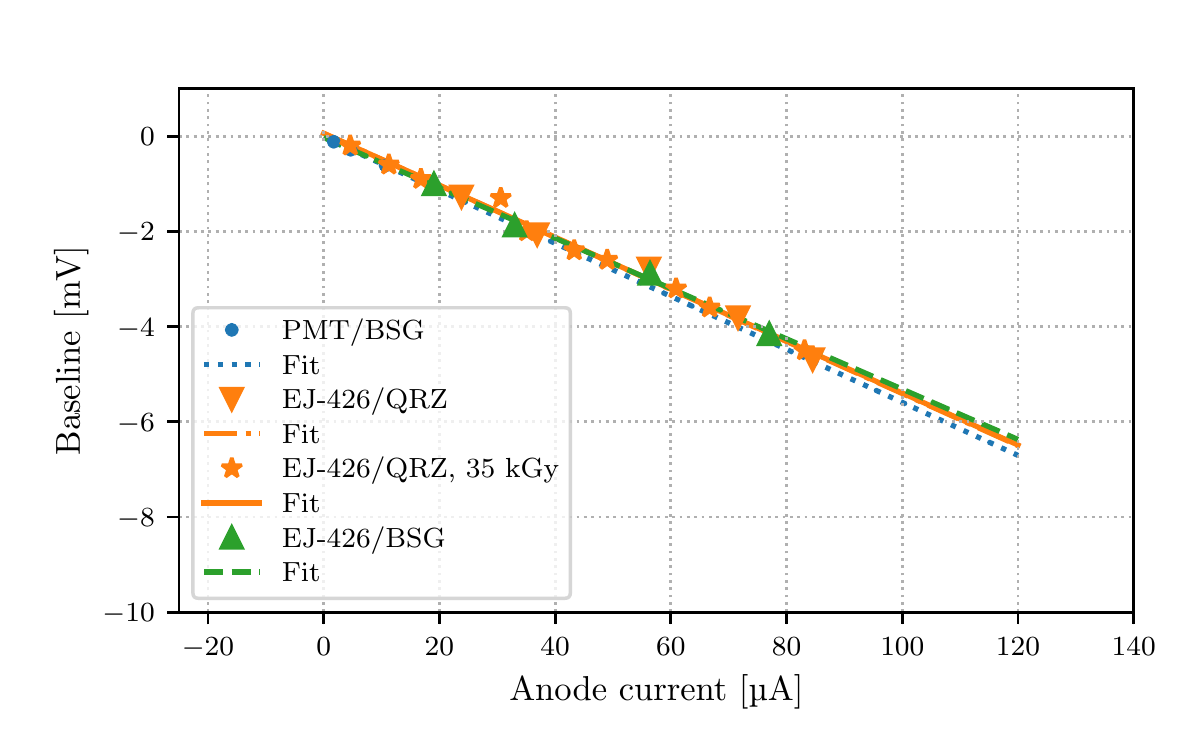}
    \caption{Signal baseline for the captured waveforms as a function of measured anode current for the different detectors. The lines indicate linear fits to the data to guide the eye. Note that the fits to the two \qrz data sets overlay each other. See text or caption of Figure~\ref{fig:CurrVsHV} for details.}
    \label{fig:BaselineVsCurr}
\end{figure}

During the measurement campaign, the anode current was no longer measured directly.
Instead, the position of the baseline in the recorded signals can be used as a proxy as Figure~\ref{fig:BaselineVsCurr} demonstrates:
Shown are the linear relationships between the baseline for the waveforms captured with the LeCroy oscilloscope and the measured anode current.
These relationships were used in post-installation to deduce, monitor and limit the anode current when, for example, deciding on an increase to the high voltage to compensate for detector degradation.
The same correlation holds true after the irradiation as indicated by the data for the \qrz detector.

\begin{figure}
    \centering
    \includegraphics[width=\textwidth]{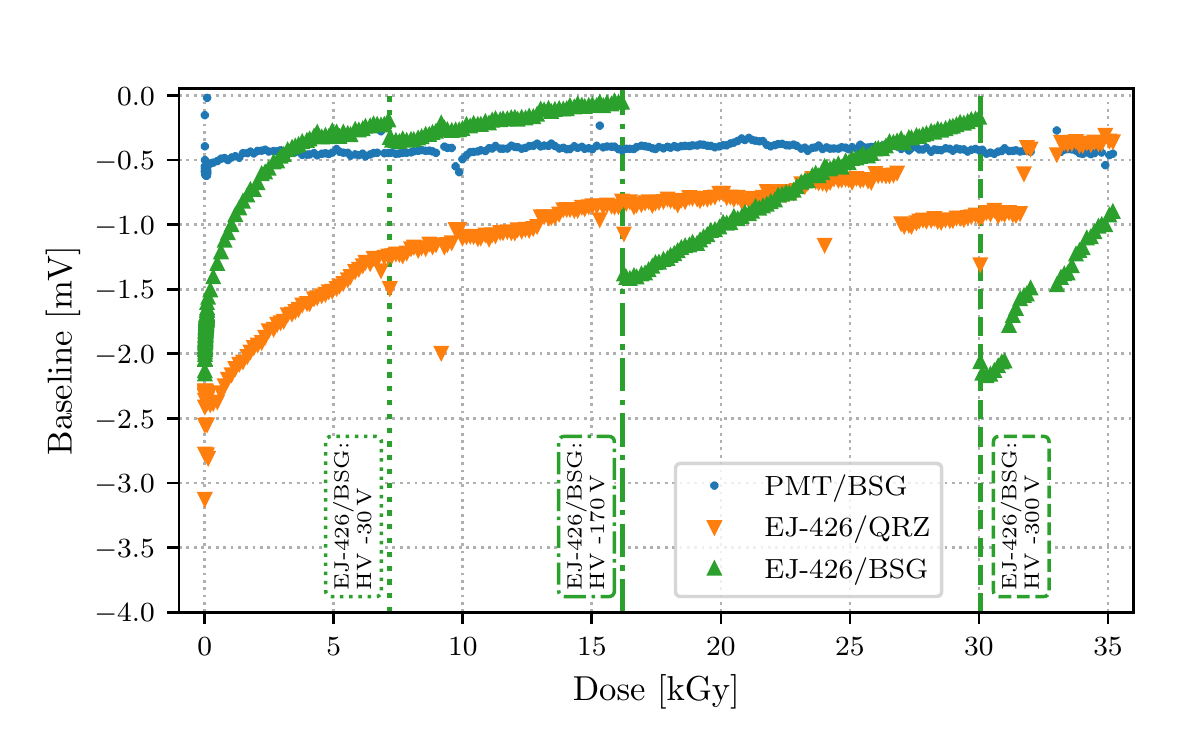}
    \caption{Baseline as a function of dose for the \bare (blue points), the \qrz (downwards-pointing orange triangles) and the \bsg (upwards-pointing green triangles). The vertical shifts for the \bsg coincide with increases of the applied high voltage by \SI{-30}{\volt}, \SI{-170}{\volt} and \SI{-300}{\volt} as indicated by the green vertical lines shown as dotted, dash-dotted and dashed, respectively. (For interpretation of the references to color in this figure caption, the reader is referred to the web version of this article.)}
    \label{fig:BaselineVsDose}
\end{figure}

Figure~\ref{fig:BaselineVsDose} shows the result of the continuous monitoring of the baseline for the different detectors as a function of the dose.
A clear shift to lower (more negative) baseline values due to the initial exposure to the high gamma and neutron flux at the measurement position was observed for all detectors.
As a reference, the baseline for all detectors was measured to be \SI{0.0}{\milli\volt} at a position with low radiation levels.
At around \SI{-0.5}{\milli\volt}, the baseline shift for the \bare detector is rather small and constant over the investigated dose range.

The most significant change in baseline value with dose is observed with the \bsg detector.
From an initial value of \SI{-2}{\milli\volt}, the shift decreases to around \SI{-0.3}{\milli\volt} at a dose of \SI{5}{\kilo\gray}.
This is probably due to a combination of browning of the PMT window and decrease in phosphorescent light from the scintillator making the detector less sensitive to the continuous gamma-ray interactions.
Three times, corresponding to doses of approx. \SI{7}{\kilo\gray}, \SI{16}{\kilo\gray} and \SI{30}{\kilo\gray}, the high voltage of the \bsg detector was increased in subsequent steps of \SI{-30}{\volt}, \SI{-170}{\volt} and \SI{-300}{\volt}, respectively.
This resulted in noticeable rises in the baseline shift indicating that some of the loss of sensitivity due to degradation effects might be recovered by increasing the gain of the PMT.

For the \qrz detector the initial baseline shift decreases more slowly over the investigated dose range.
It finally reaches a value of \SI{-0.5}{\milli\volt} at a dose of $\approx \SI{35}{\kilo\gray}$ from an initial value of \SI{-3.0}{\milli\volt}.
As the two latter detectors differ mainly in the PMT window material, likely browning of the PMT window has the biggest effect on the performance of the \bsgns.

Single points in Figure~\ref{fig:BaselineVsDose} that are shifted by an offset to the global trend can be traced back to restarts of the DAQ system.
The cause for the drop of the baseline values for the \qrz detector between $\SIrange{27}{32}{\kilo\gray}$ of \SI{0.4}{\milli\volt}, however, is not known.

\subsection{Identified neutrons as function of dose}
\label{sec:neutronsVSdose}

\begin{figure}
    \centering
    \includegraphics[width=\textwidth]{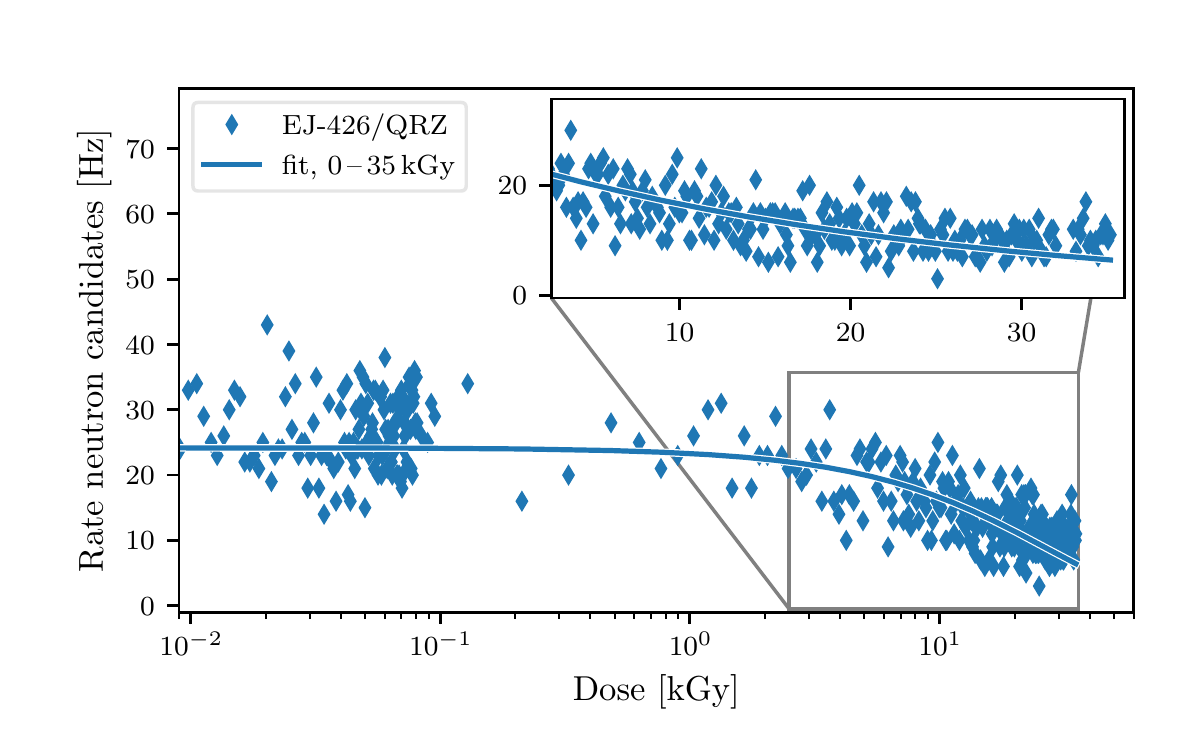}
    \caption{Number of identified neutron candidates as a function of dose recorded with the \qrz detector. The solid line shows an exponential fit to the data to guide the eye. The inset highlights the dose region $\SIrange{2.5}{35}{\kilo\gray}$ on a linear scale.}
    \label{fig:ncandidatesSilica}
\end{figure}

\begin{figure}
    \centering
    \includegraphics[width=\textwidth]{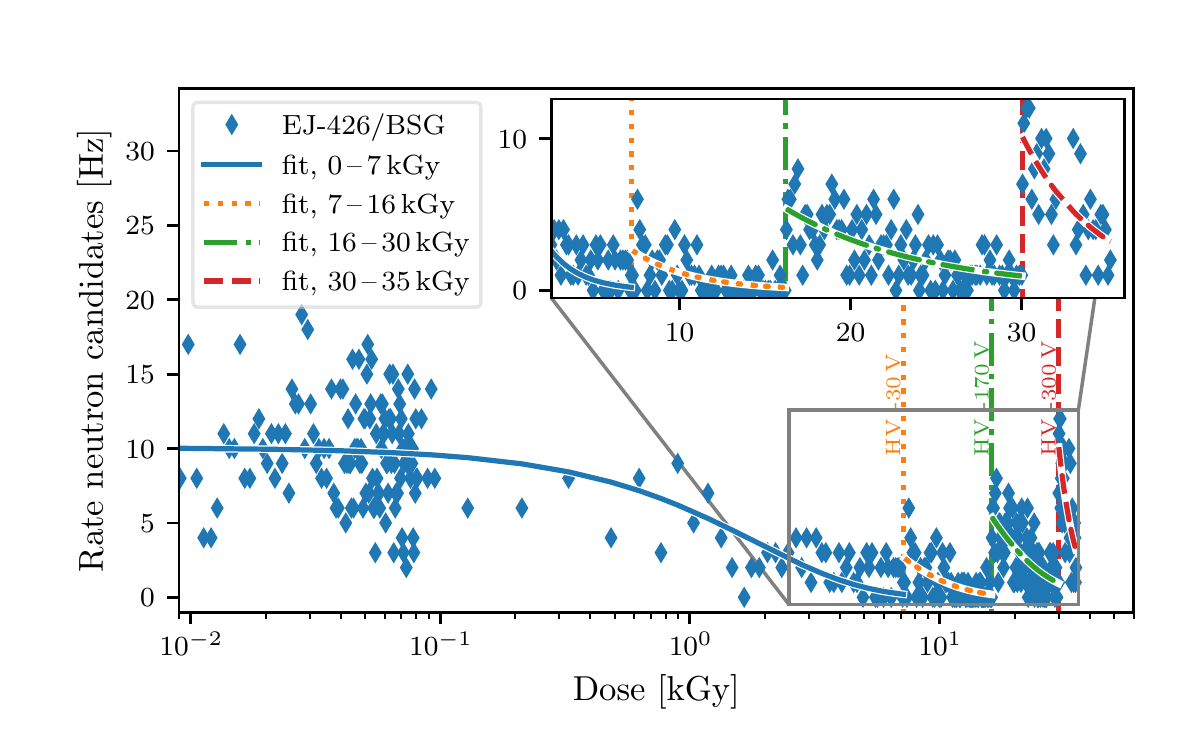}
    \caption{Number of identified neutron candidates as a function of dose recorded with the \bsg detector.
      The vertical lines shown in dotted orange, dash-dotted green and dashed red indicate increases of the applied high voltage by \SI{-30}{\volt}, \SI{-170}{\volt} and \SI{-300}{\volt}, respectively.
      The functions shown as lines are exponential fits to the corresponding data sections to guide the eye. The inset highlights the dose region $\SIrange{2.5}{35}{\kilo\gray}$ on a linear scale.
      (For interpretation of the references to color in this figure caption, the reader is referred to the web version of this article.)}
    \label{fig:ncandidatesBorosilicate}
\end{figure}

Figures~\ref{fig:ncandidatesSilica} and \ref{fig:ncandidatesBorosilicate} show the number of identified neutron candidates in the acquired signals as a function of the dose for the \qrz and \bsg detectors, respectively.
The main figures show the entire data set on a logarithmic scale between $\SIrange{1e-2}{40}{\kilo\gray}$ while the insets highlight the region $\SIrange{2.5}{35}{\kilo\gray}$ on a linear scale.
The lines in the figure indicate fits of the exponential decay function to the data, $N(d) = N_0 e^{-\frac{1}{\tau} d}$, where $d$ corresponds to the dose, $\tau$ to the exponential time constant, and $N_0$ to the number of initial neutron candidates.
For the \bsg detector, the steps in high voltage are indicated by vertical lines which separate the regions used in the fits.

As each point corresponds to a single acquisition of \SI{1}{\second} of data, the resulting number of neutron candidates per second fluctuates strongly between measurements.
For the \qrzns, the initial frequency of detected neutrons is between $\SIrange{20}{35}{\per\second}$ while for the \bsg it varies between $\SIrange{4}{20}{\per\second}$.
Over the investigated dose range, however, the number of neutron candidates is clearly decreasing as function of the dose for both detectors.

For the \qrz detector, the exponential time constant was determined to $\tau_\mathrm{\qrz} = \SI{27}{\kilo\gray}$, whereas for the \bsg detector the initial decrease is much more rapid with $\tau_\mathrm{\bsg} = \SI{2}{\kilo\gray}$.
For the former, this corresponds to an identification of \sfrac{1}{3} of the initial number of identified neutron candidates at the maximum obtained dose.
For the latter, an increase and subsequent decrease in number of identified neutrons can be seen at the three occasions when the high voltage to the detector had been increased.
These results therefore follow a similar trend as seen for the baseline shift in Figure~\ref{fig:BaselineVsDose}.

Over the whole measurement period, no neutron candidates were found in the signal from the \bare detector indicating that the identification algorithm is not sensitive to the noise seen in a bare PMT.

The total accumulated neutron fluence was verified by comparing it against the activation level of the tantalum foil which was recovered from the detector vessel when the measurement campaign was concluded.
The activity of the foil was measured in the laboratory to be \SI{225}{\Bq}.
This corresponds to an average neutron flux of 360 neutrons per second through the foil, in the energy interval for the neutron capture reaction of $^{181}$Ta, as described in Section~\ref{sec:neutronDose}.
However, due to the simplified model used in the simulations, it should be noted that this value is a best-guess estimate and is associated with a significant systematic uncertainty.
Using the energy distribution obtained from simulations, the EJ-426 is expected to have an neutron capture efficiency in $^{6}$Li of 15\%.
From this value, the average flux through the EJ-426 scintillators can be estimated to 280 neutrons per second of which 42 neutrons per second are expected to react and to leave a detectable signal.
This implies an accumulated neutron fluence in the order of $\SI{5e9}{\n\per\cm\squared}$.

In the beginning of the campaign the \qrz measured on average $\sim27$ neutrons per second (see Figure~\ref{fig:ncandidatesSilica}) and in the end approximately 11 neutrons per second.
This corresponds to an estimated detection efficiency of 60\% of the expected neutron captures in $^{6}$Li initially and 25\% towards the end of the campaign.

\begin{figure}
    \centering
    \includegraphics[width=\textwidth]{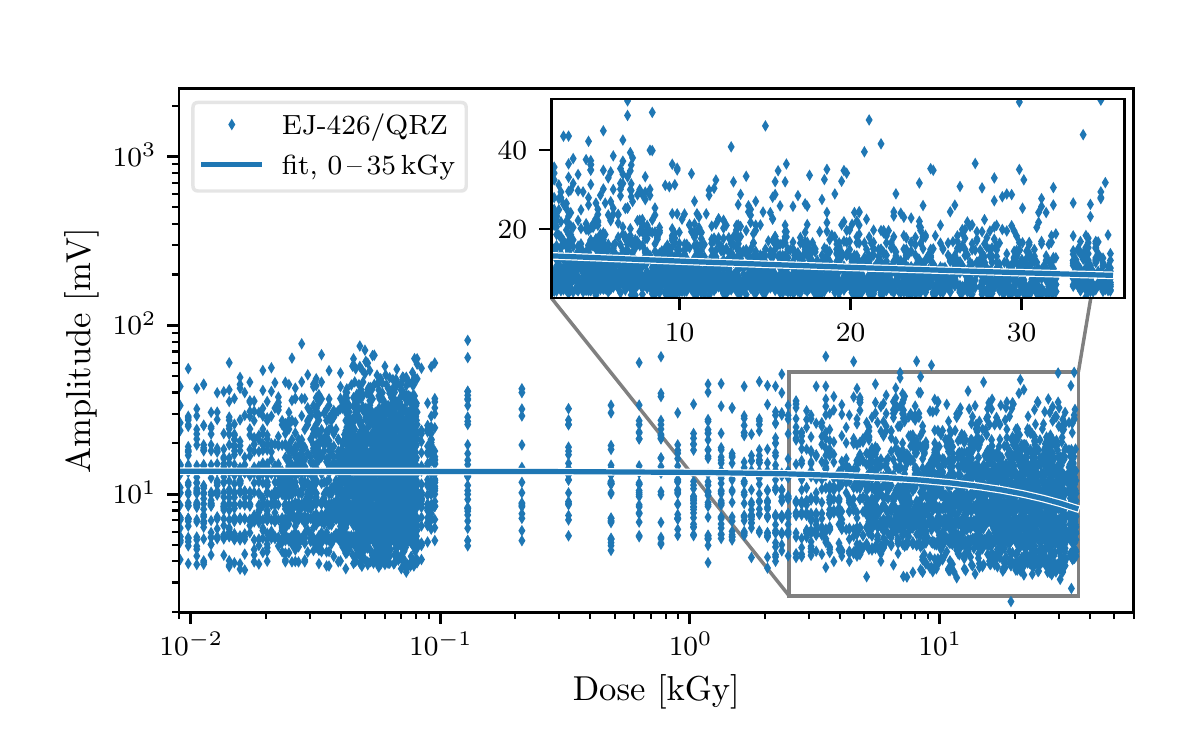}
    \caption{Amplitude values of identified neutron candidates for the \qrz detector as function of dose on a double-logarithmic scale. A fitted exponential curve is included to guide the eye. The inset highlights the dose region $\SIrange{2.5}{35}{\kilo\gray}$ on a linear scale. See text for details.}
    \label{fig:ampQRZ}
\end{figure}

\begin{figure}
    \centering
    \includegraphics[width=\textwidth]{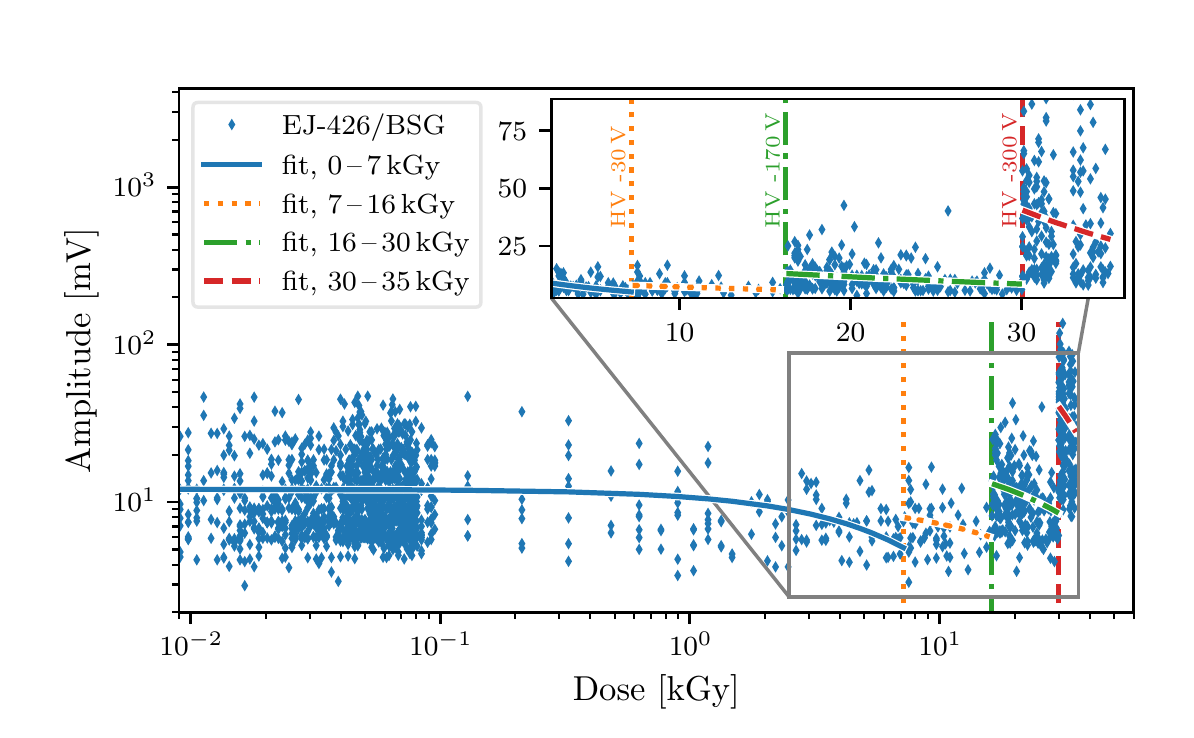}
    \caption{Amplitude values of identified neutron candidates for the \bsg detector as function of dose on a double-logarithmic scale.
      The vertical lines shown in dotted orange, dash-dotted green and dashed red indicate increases of the applied high voltage by \SI{-30}{\volt}, \SI{-170}{\volt} and \SI{-300}{\volt}, respectively.
      The functions shown as lines are exponential fits to the corresponding data sections to guide the eye.
      The inset highlights the dose region $\SIrange{2.5}{35}{\kilo\gray}$ on a linear scale.
      (For interpretation of the references to color in this figure caption, the reader is referred to the web version of this article.)}
    \label{fig:ampBSG}
\end{figure}

Figures~\ref{fig:ampQRZ} and \ref{fig:ampBSG} show measured pulse amplitudes of each identified neutron candidate for the \qrz and \bsg detectors, respectively.
The amplitude values have been baseline-corrected and converted from units of ADC into Voltages using the prescription provided by the manufacturer of the HS6 digitizer.
The main figures show the entire data set on a double-logarithmic scale between $\SIrange{1e-2}{40}{\kilo\gray}$ while the insets highlight the region $\SIrange{2.5}{35}{\kilo\gray}$ on a linear scale.
Non-vertical lines in the figure indicate fits of the exponential decay function to the data.
For the \bsg detector, the steps in high voltage are again indicated by vertical lines which separate the regions used in the corresponding fit.

Initially, the values of the pulse amplitudes are spread between $\SIrange{4}{70}{\milli\volt}$ and $\SIrange{4}{50}{\milli\volt}$ for the \qrz and \bsgns, respectively.
A decrease in maximum and mean amplitude (indicated by the fit) are seen in both detectors as the accumulated dose increases.
The minimum amplitude, however, is stable around \SI{4}{\milli\volt} and reflects the threshold inherent to the neutron identification algorithm described in Section~\ref{sec:data-sampl-analys}.

As previously noted, the performance degradation seen in the \bsg detector can be partially recovered by increases of the high voltage to the detector.
This suggests that the observed loss of the number of neutron candidates is due to an overall reduction of signal amplitude: with accumulated dose an increasingly larger fraction of the neutron pulses fall below the threshold of the system.
Likely causes are a degradation of the optical transparency of the scintillator material as well as the PMT window combined with a reduction of PMT gain due to both the accumulated dose and the high average anode current.
The capture probability of neutrons in the EJ-426 scintillator material itself seems unaffected.
This would be consistent with the observations by~\cite{HamamatsuPMTHandbook4E} which have reported a similar degradation of the optical properties at comparable gamma-ray doses.
Radiation damage effects from neutrons, on the other hand, are only expected at significantly higher doses~\cite{iida90_effec_neutr_irrad_optic_compon_fusion_diagn}.

It should be noted that increases in high voltage also give rise to noise seen on top of the neutron pulses.
This complicates the reliable identification in particular of low-amplitude pulses close to the noise threshold.
Any significant change in the high voltage might therefore require adjustments to the parameters of the presented neutron identification algorithm in order to sustain its efficiency and error-rate.

\subsection{Analysis of radiation-induced degradation after the concluded measurement campaign}

\begin{figure}[htbp]
  \centering
  \includegraphics[width=12cm]{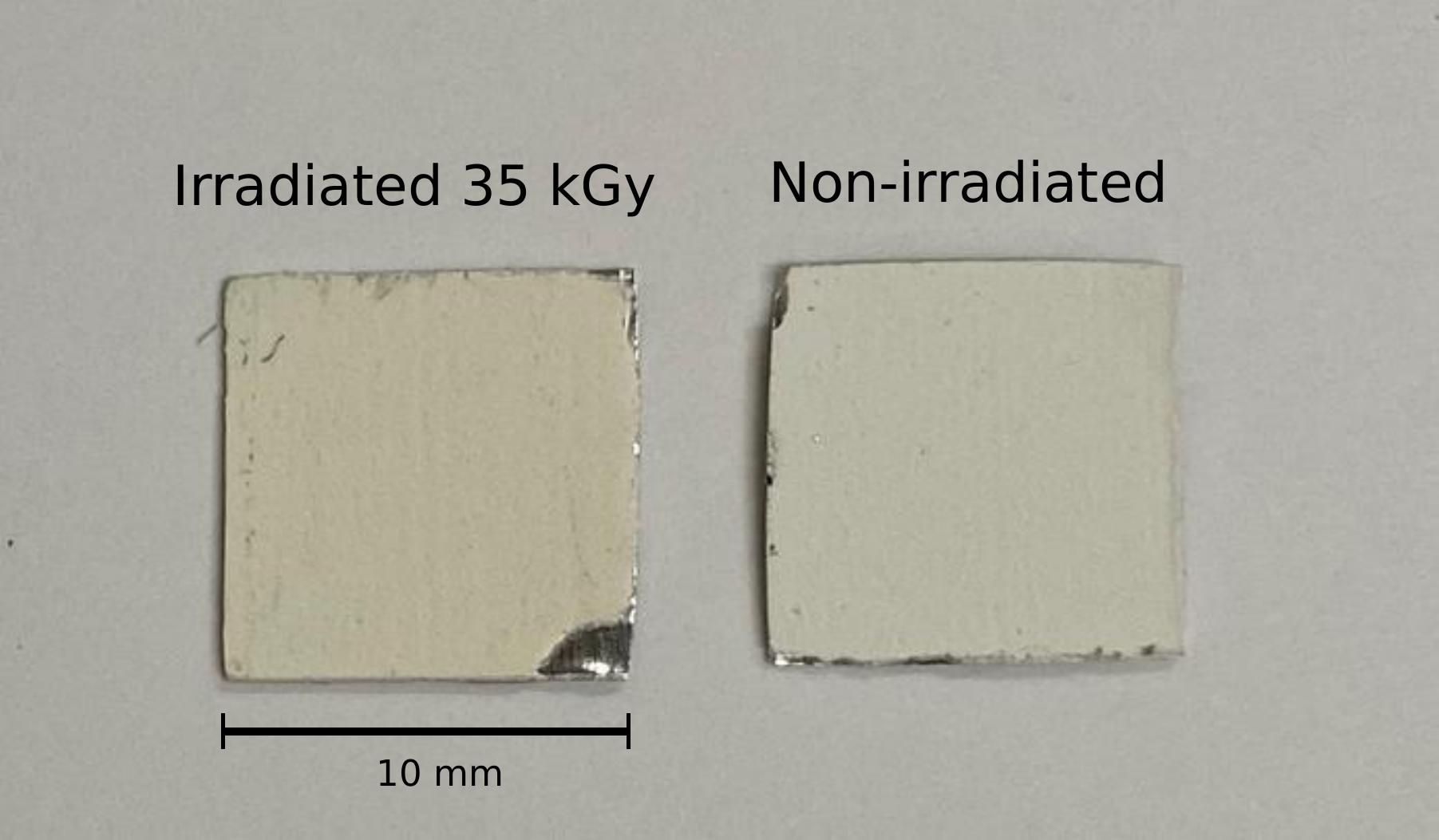}
  \caption{Left: Photographs of EJ-426 scintillator screens. Left: irradiated with approximately \SI{35}{\kilo\gray} as part of the \bsg assembly. Right: non-irradiated.}
  \label{fig:IrrEj426}
\end{figure}

\begin{figure}[htbp]
  \centering
  \includegraphics[width=12cm]{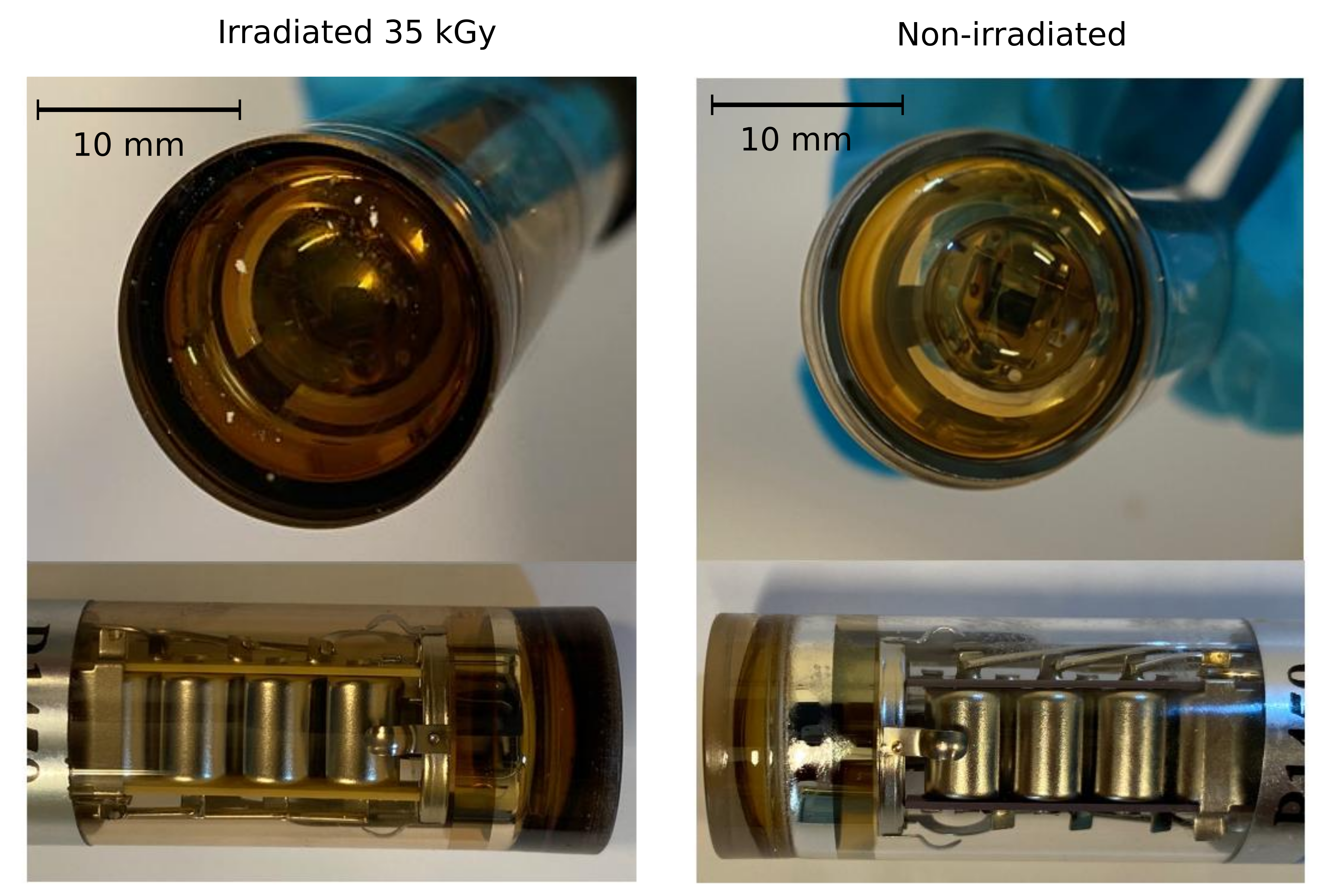}
  \caption{Photographs of two Hamamatsu R1450 photo multiplier tubes. Left: irradiated with approximately \SI{35}{\kilo\gray}. Right: non-irradiated.}
  \label{fig:IrrPMT}
\end{figure}

After the measurement campaign the tantalum foil, the \bare and \bsg detectors were recovered and analyzed in the laboratory.
Figure~\ref{fig:IrrEj426} shows a photograph of the irradiated EJ-426 scintillator screen from the \bsg next to a non-irradiated one.
Some coloring of the irradiated screen can be noted.
Figure~\ref{fig:IrrPMT} shows a photograph of the irradiated Hamamatsu R1450 PMT next to an non-irradiated PMT.
Clear browning of the entrance window as a result of the radiation can be seen.

\begin{figure}[htbp]
  \centering
  \includegraphics[width=\textwidth]{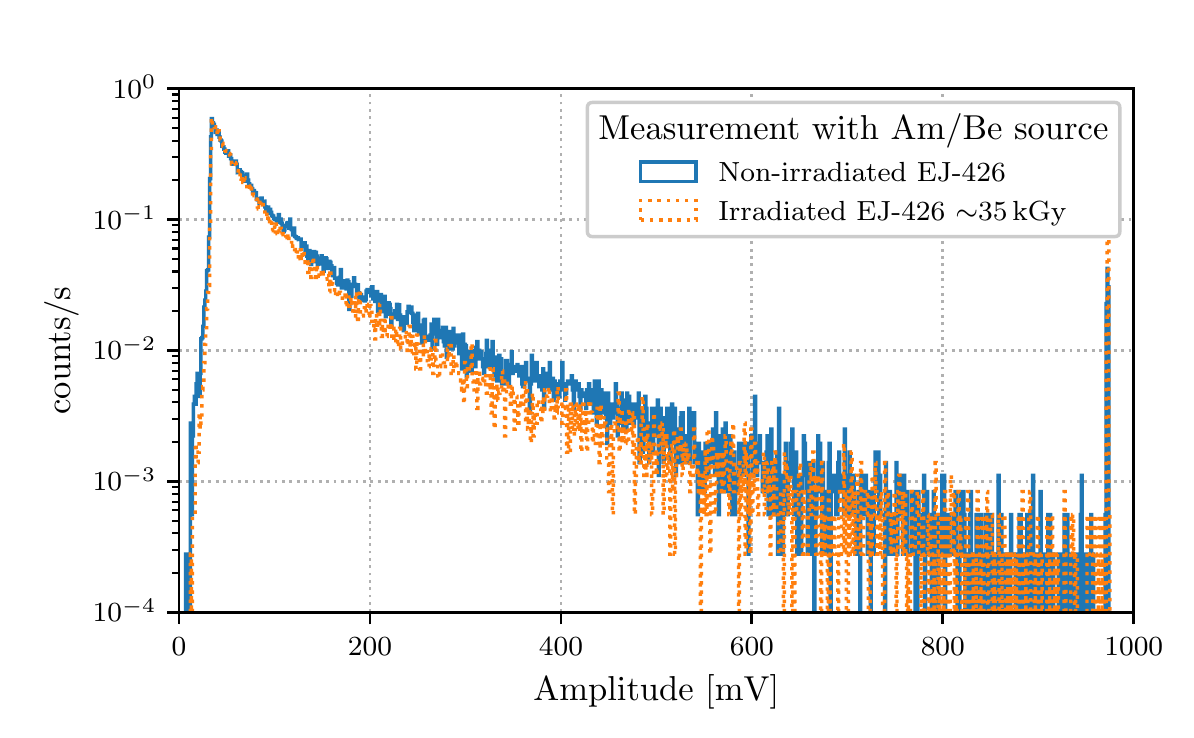}
  \caption{Signal amplitude spectra from an Americium/Beryllium source measured with an non-irradiated (solid blue histogram) and irradiated (dotted orange histogram) EJ-426 scintillator screen, each mounted on the same non-irradiated Hamamatsu H1949-51 PMT. (For interpretation of the references to color in this figure caption, the reader is referred to the web version of this article.)}
  \label{fig:IrrEj426Amp}
\end{figure}

The scintillator screens shown in Figure~\ref{fig:IrrEj426} were mounted on a new Hamamatsu H1949-51 PMT and exposed to moderated neutrons from an Americium/Beryllium source.
Figure~\ref{fig:IrrEj426Amp} shows the signal amplitude distributions for the detectors recorded during one hour of measurement.
Both detectors exhibit very similar spectra.
While the average amplitude for the detector with the irradiated screen is somewhat shifted towards lower energies compared to the non-irradiated sample, the maximum signal amplitude seems to be largely unchanged.
This small change in the distribution could be coupled to the coloring of the irradiated EJ-426 screen which reduced its transparency to its own scintillation light.
However, the shift could also be a result of, for example, the coupling between the screen and the PMT window.

\begin{figure}[htbp]
  \centering
  \includegraphics[width=\textwidth]{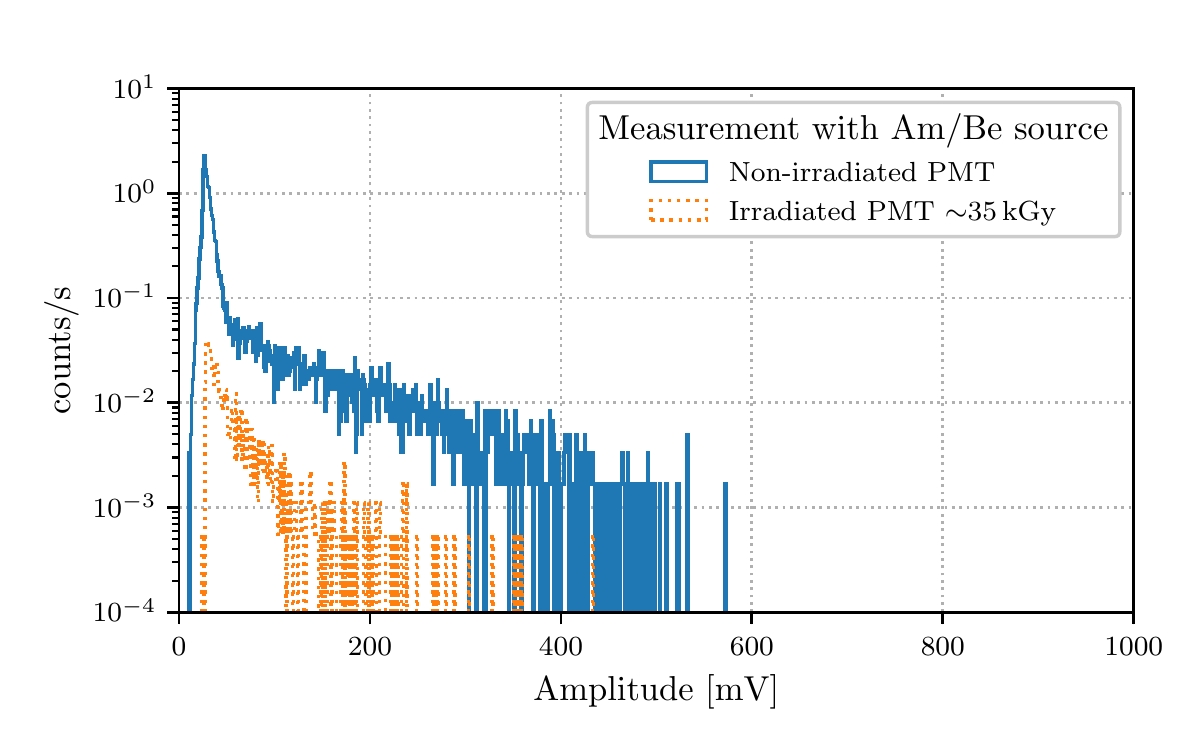}
  \caption{Signal amplitude spectra from an Americium/Beryllium source measured with an non-irradiated EJ-426 scintillator screen mounted a non-irradiated (solid blue histogram) and irradiated (dotted orange histogram) Hamamatsu R1450 PMT with BSG window. The latter was retrieved from the \bsg detector. (For interpretation of the references to color in this figure caption, the reader is referred to the web version of this article.)}
  \label{fig:IrrPMTAmp}
\end{figure}

Figure~\ref{fig:IrrPMTAmp} compares the pulse amplitudes for the PMT of the \bsg detector when exposed to moderated neutrons from the Americium-Beryllium source before and after irradiation.
The blue line had been recorded before the measurement campaign while the data of the orange line was taken after the irradiation but with a non-irradiated sheet of EJ-426 scintillator mounted on the irradiated PMT.
The two distributions show a much more pronounced difference than seen in the scintillator screen comparison of Figure~\ref{fig:IrrEj426Amp}.
The sprectrum from the irradiated detector is shifted to lower values resulting in fewer total counts rate over the threshold.
This is likely due to the browning of the entrance window of the PMT as well as gain loss due to the accumulated dose and the high and prolonged anode current.

\section{Summary and Conclusions}

Neutron detectors were operated in a harsh radiation environment close to spent nuclear fuel elements and their performance monitored in-situ over several months.
Both the \qrz and the \bsg detectors were able to measure neutrons despite the intense gamma-ray background dose rate of approximately \SI{6}{Sv/h}.
During the measurement campaign, the detectors accumulated a total dose of \SI{35}{\kilo\gray}.

Over this dose range, both detectors show a performance degradation and a reduction of the number of neutrons detected.
As seen from the tests conducted on the \bsg detector the neutron detection efficiency can be partially restored by increasing the high voltage of the detector.
Without adjustment, about 30 \% of the initial number of neutrons could be detected with the \qrz detector at the end of the measurement campaign.
The base line of the detector signals can be easily monitored and can serve as an online indication for the degradation of the detector sensitivity provided that the gamma-ray background is comparable.

A loss of signal amplitude has been identified as main reason for the reduced number of neutron counts seen as a result of the irradiation.
Browning of the PMT window and a loss of gain have been found to be the main causes in case of the \bsg detector.
As expected, the window material of the \qrz has been shown to be significantly more radiation hard than that of the \bsgns.
For the \qrz detector the observed degradation of performance is likely due to gain loss in the PMT caused by the high average anode current.

Particularly low-amplitude pulses are challenging to identify algorithmically as they are close to the noise threshold.
Optimizations of the algorithm could improve these results further.
Alternatively, promising outcomes were archived in an early study applying machine learning techniques to the data.

For this work, an efficient trigger-less data acquisition system has been evaluated with good results.
Despite the limited access on-site, the up-time of the detector system has been more than 95\% and the system has been remotely controlled only via exchanges of an attached USB thumb drive.

This first study of the performance of non-$^{3}$He neutron detectors based on commercially-available components showed that the evaluated detector technology is capable of measuring neutrons in close proximity to spent nuclear fuel and in the associated high-intensity mixed-field radiation environment even after prolonged exposure.
The accumulated dose of \SI{35}{\kilo\gray} that the investigated detectors were exposed to corresponds to several years of use in a future detector system at the foreseen encapsulation plant.

\section*{Acknowledgments}
The authors would like to thank Andreas Högström at CLAB for helping with the installation of the measurement system as well as exchanging and returning the thumb drives.

The commissioning of the detector system and analysis of the irradiated components were carried out at the Source Testing Facility and Applied Nuclear Physics Laboratory, Division of Nuclear Physics, Lund University.

We acknowledge the support by the Swedish Nuclear Fuel and Waste Management Company (SKB).


\bibliography{references}

\end{document}